\documentclass[10pt,letterpaper]{article}
\usepackage[utf8x]{inputenc}
\usepackage[table]{xcolor}
\usepackage{amsmath}
\usepackage{amssymb}
\usepackage{cite}
\usepackage{hyperref}
\usepackage{nameref}
\usepackage{url}
\usepackage{array}
\usepackage{fullpage}
\usepackage{graphicx}
\usepackage{multirow}
\usepackage{authblk}
\usepackage{tabularx}
\usepackage{titlesec}

\title{Fast frequency modulation is encoded according to the listener expectations in the human subcortical auditory pathway}
\author[12]{Alejandro Tabas}
\author[2]{Stefan Kiebel}
\author[34]{Michael Marxen}
\author[12]{Katharina von Kriegstein}
\affil[1]{Department of Psychology, Technische Universit\"{a}t Dresden, Dresden, Germany}
\affil[2]{Max Planck Institute for Human Cognitive and Brain Sciences, Leipzig, Germany}
\affil[3]{Department of Psychiatry, Technische Universit\"{a}t Dresden, Dresden, Germany}
\affil[4]{Neuroimaging Center, Technische Universit\"{a}t Dresden, Dresden, Germany}
\date{}

\titleformat*{\section}{\large\bfseries}
\titleformat*{\subsection}{\bfseries}

\begin{document}

  \maketitle

  \abstract{

    Expectations aid and bias our perception. In speech, expected words are easier to recognise than unexpected words, particularly in noisy environments, and incorrect expectations can make us misunderstand our conversational partner. Expectations are combined with the output from the sensory pathways to form representations of speech in the cerebral cortex. However, it is unclear whether expectations are propagated further down to subcortical structures to aid the encoding of the basic dynamic constituent of speech: fast frequency-modulation (FM). Fast FM-sweeps are the basic invariant constituent of consonants, and their correct encoding is fundamental for speech recognition. Here we tested the hypothesis that subjective expectations drive the encoding of fast FM-sweeps characteristic of speech in the human subcortical auditory pathway. We used fMRI to measure neural responses in the human auditory midbrain (inferior colliculus) and thalamus (medial geniculate body). Participants listened to sequences of FM-sweeps for which they held different expectations based on the task instructions. We found robust evidence that the responses in auditory midbrain and thalamus encode the difference between the acoustic input and the subjective expectations of the listener. The results indicate that FM-sweeps are already encoded at the level of the human auditory midbrain and that encoding is mainly driven by subjective expectations. We conclude that the subcortical auditory pathway is integrated in the cortical network of predictive speech processing and that expectations are used to optimise the encoding of even the most basic acoustic constituents of speech.
  
  }

  \section*{Introduction}

    Expectations can have dramatic effects on sensory processing \cite{deLange2018}. A prime example is speech perception, where expectations influence processing on many different levels. For instance, word recognition is often completed before the entire word has been heard \cite{Marslen1973}, and strongly affected by semantic context \cite{Davis2011}, word prevalence \cite{Sereno2003}, and prior knowledge \cite{Sohoglu2012}.

    Predictive coding is one of the leading frameworks explaining how expectations affect perceptual encoding \cite{Rao1999, Friston2003, Friston2005}. Predictive coding has been suggested to form the basis for encoding of speech in the auditory system \cite{Sohoglu2012, Pickering2013, Kuperberg2016, Cope2017, Friston2021}. Predictive coding formulates perception as a constant process of hypothesis testing \cite{Friston2003, Friston2005}: neural centres at high-order levels perform predictions on the sensory world via a generative model, while neurons at lower levels test these predictions against the actual sensory input \cite{Spratling2017, Keller2018}. A key hypothesis of the framework is that sensory neurons at lower levels do not encode the features of the stimuli but prediction error: the difference between the sensory input and the expectations of the observer on that sensory input. When the input matches the expectations, prediction error neurons do not need to communicate with higher-level centres, speeding up recognition and optimising the neural code. 

    In the human auditory cortex, speech sounds are encoded as prediction error \cite{Blank2016, Ylinen2016, Blank2018, Vidal2019, Signoret2020, Sohoglu2020, Hovsepyan2020}; whether that is also the case in the subcortical auditory pathway is unclear. Anatomical and physiological properties make the subcortical auditory pathway a prime candidate for predictive coding: The subcortical auditory pathway contains massive cortico-thalamic and cortico-collicular efferent systems \cite{Lee2011, Schofield2011, Winer2005b, Winer1984} that are well suited to transmit complex expectations to subcortical nuclei. Moreover, neural populations in the subcortical auditory pathway are endowed with much shorter time constants and faster access to acoustic information than neural populations in the cerebral cortex \cite{Steadman2018}. This feature renders subcortical auditory pathways better suited to test hypotheses on the incoming fast dynamics of speech sounds \cite{Giraud2000, Kriegstein2008, Osman2018}.

    Stimulus-specific adaptation (SSA) has been used as a first attempt to test for predictive coding in the subcortical pathways. SSA is a phenomenon where individual neurons adapt to repetitions of a pure tone but show recovered responses to a frequency deviant \cite{Ulanovsky2003}. SSA is present in single neurons of the rodent's auditory thalamus (medial geniculate body; MGB) \cite{Parras2017, Bauerle2011, Antunes2010, Anderson2009} and auditory midbrain (inferior colliculus; IC) \cite{Parras2017, Robinson2016, Ayala2015, Gao2014, Perez2012, Zhao2011}, and in neural populations of the human IC and MGB \cite{Tabas2020, Cacciaglia2015, Cornella2015, Escera2014, Gao2014, Grimm2011}. SSA can, however, be explained both by neural habituation or predictive coding (see~\cite{Tabas2021} for a review). In the case of pure tones, we have recently used a novel SSA paradigm which revealed that SSA in human IC and MGB is driven largely by subjective expectations of the listeners, as hypothesised by predictive coding but not by neural habituation \cite{Tabas2020}.

    In contrast to pure tones, speech sounds comprise highly dynamic elements. These elements cannot be fully characterised by mixtures of static pure tones. The most ubiquitous of these dynamic elements are fast frequency-modulated (FM)-sweeps \cite{Liberman1956, Liberman1978}. Combinations of three fast FM-sweeps of different average frequency acoustically characterise consonants preceding a vowel. Accurately perceiving the modulation direction and rate of those FM-sweeps is crucial for speech comprehension; for instance, the phonemes /ba/ and /da/ differ only on the modulation direction of one of their comprised FM-sweeps \cite{Liberman1956}.

    Whilst pure tones are encoded according to their frequency along the tonotopic axis already at the basilar membrane \cite{Hu2003}, FM-sweeps are encoded in FM-direction and FM-rate selective neurons \cite{Kuo2012}. In humans, the lowest level in the auditory hierarchy with evidence for fast FM-direction \cite{Hsieh2012, Joanisse2014} and rate \cite{Okamoto2015} selectivity is in auditory cortex. Although FM-sensitive neurons have been reported in the rodent IC and MGB \cite{Zhang2003, Lui2003, Ye2010, Kuo2012, Issa2016}, it is currently unclear whether FM is also encoded in subcortical stations of the human auditory pathway. If that was the case, an important question is whether FM-sweeps, as basic constituent of speech sounds, are encoded according to the subjective expectations of the listener.

    Given the paramount importance of fast FM-sweeps for speech comprehension and the privileged temporal properties of the subcortical elements of the auditory pathway, we addressed two key questions. First, whether FM-rate and FM-direction are already encoded in neural populations of the human IC and MGB. Second, whether fast FM-sweeps are encoded in IC and MGB according to the principles of predictive coding; i.e., as prediction error with respect to the subjective expectations of the listener. To address these questions we measured blood-oxygen level dependent (BOLD) responses in two key subcortical structures of the auditory pathway, IC and MGB, while participants listened to sequences of FM-sweeps. The FM-sweeps were constructed as closely as possible to FM-sweeps characteristic of speech sounds, and designed in such a way that they all elicited the same pitch percept and the same average activity along the tonotopic axis \cite{Nabelek1970, Tabas2021b}. To test if encoding was mediated by predictive coding, we introduced abstract rules that assigned different likelihoods to different FM-sweeps without affecting stimulus statistics \cite{Tabas2020}. We reasoned that, if FM-sweeps were encoded according to their objective properties, an FM-sweep embedded in a specific statistical context should elicit the same activation no matter the expectations that participants have on its occurrence. Reversely, if FM-sweeps were encoded according to the principles of predictive coding, BOLD responses should directly depend on how well the sensory input fits the expectations of the listeners.

    Once established that fast FM-sweeps are encoded as prediction error in IC and MGB, we addressed two additional questions. First, we investigated whether the same encoding mechanisms operate in primary (or lemniscal) and secondary (or non-lemniscal) MGB. Previous findings in the animal literature indicated that SSA and predictive coding are stronger in secondary subdivisions of the subcortical pathways (e.g.,~\cite{Parras2017}, a finding that is not apparent in humans \cite{Tabas2020, Cacciaglia2015}). Second, we tested whether the topographic distribution of neural populations encoding prediction error in IC and MGB were replicated in FM-sweeps and pure tones. Similar topographies in both stimulus families would indicate a common mechanism for subcortical predictive coding.

  \section*{Results}

    \subsection*{Experimental Design and Hypotheses}  

\begin{figure}[tbh!]
        \centering
        \includegraphics[width=0.75\textwidth]{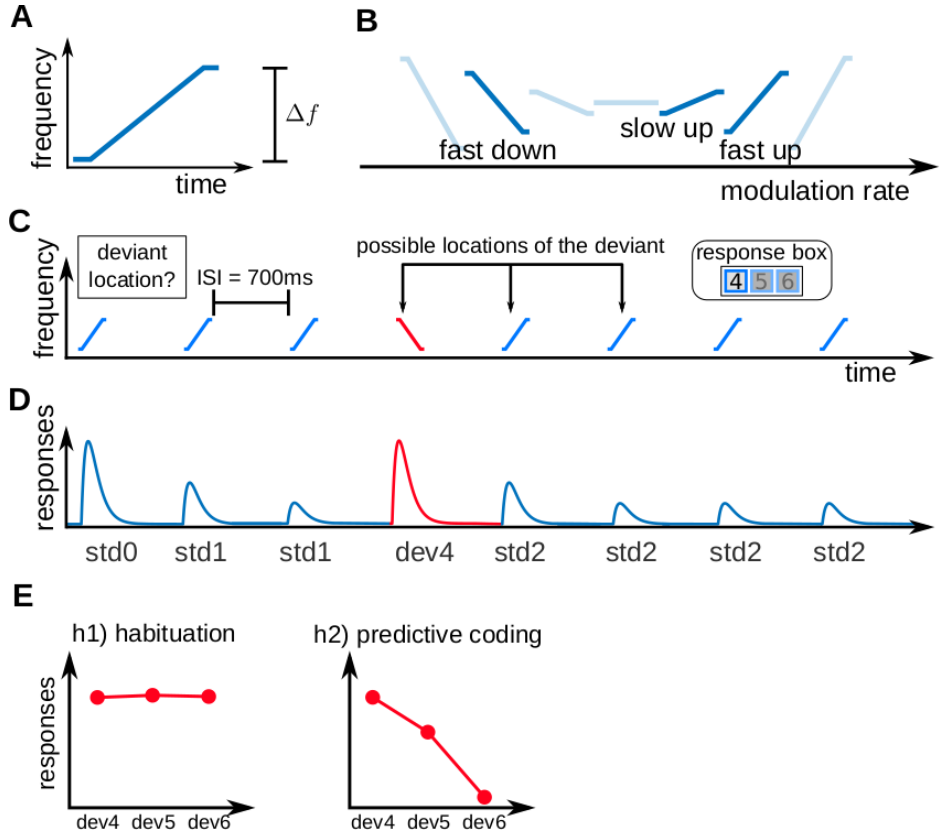}
        \vspace{1em}
        \caption{\textbf{Experimental design and hypotheses}. 
        A) Example of an FM-sweep with positive FM-rate $\Delta f$ B) The three FM-sweeps used in the experiment (in dark blue) in comparison to an hypothetical family of seven sweeps with increasing modulation rate. All sweeps had the same duration of 50\,ms. They were characterised by differences in the frequency span  $\Delta f$. C) Example trial. Each trial consisted of a sequence of seven repetitions of one FM-sweep (standards; blue) and one other FM-sweep (deviant; red). In each trial, a single deviant was located in positions 4, 5, or 6 of the sequence. Participants reported, in each trial, the position of the deviant right after they identified it. Each participant completed up to 540 trials in total, 60 per deviant position and $\Delta^2 = |\Delta f_{\text{deviant}} - \Delta f_{\text{standard}}|$. Sweeps within a sequence were separated by 700\,ms inter-stimulus-intervals (ISIs). 
        D) Schematic view of the expected underlying responses in the auditory pathway for the sequence shown in C, together with the definition of the experimental variables ($std0$: first standard; $std1$: repeated standards preceding the deviant; $std2$: standards following the deviant; $dev\,x$: deviant in position $x$). 
        E) Expected responses in the auditory pathway nuclei corresponding to the hypotheses: h1) responses reflect adaptation by habituation only; h2) responses reflect prediction error with respect to the participant's expectations.
        \label{fig:design}}
      \end{figure}

      The stimuli were three fast FM-sweeps: One sweep with a fast negative FM-rate (frequency span $\Delta f = -200$\,Hz), one with a fast positive FM-rate (frequency span $\Delta f = 200$\,Hz), and one with a slow positive FM-rate (frequency span $\Delta f = 100$\,Hz; Fig~\ref{fig:design}A-B). We used 50\,ms long sweeps in the frequency range of $f\sim 1500\,$Hz so that they had the typical properties of formant transitions in speech \cite{Liberman1978}. The sweep average frequencies were adjusted so that all FM-sweeps elicited the same average activity along the tonotopic axis and were perceived as having the same pitch \cite{Tabas2021b, Nabelek1970}; this design guaranteed that FM-direction and FM-rate selective neurons were necessary to differentiate between any two sweeps in the paradigm. 
          
      We arranged the stimuli in sequences of 8 FM-sweeps with 7 repetitions of the same sweep (standard) and one deviating sweep (deviant) (Fig~\ref{fig:design}C). Participants were instructed to report, with a button press, the position of the deviant within the sequence as fast and accurately as possible after identifying the deviant. Each sequence was characterised by the position of the deviant and $\Delta^2 f = |\Delta f_{\text{deviant}} - \Delta f_{\text{standard}}|$: the absolute difference between the frequency spans of the deviant and the standard. With the three FM-sweeps we built three different combinations: one with $\Delta^2 f = 100$\,Hz, where the two FM-sweeps have the same direction (up) but different modulation rate; one with $\Delta^2 f = 300$\,Hz, where the two FM-sweeps have different rate and different direction; and one with $\Delta^2 f = 400\,$Hz, where to two FM-sweeps have the same modulation rate ($|\Delta f| = 200$\,Hz) but different direction (Fig~\ref{fig:design}B). 

      Expectations for each of the deviant positions were modulated by two abstract rules that were disclosed to the participants: 1) all sequences have a deviant, and 2) the deviant is always located in position 4, 5, or 6. The three deviant positions were used the same number of times along the experiment, so that the three deviant positions were equally likely at the beginning of the sequence. Therefore, the likelihood of finding a deviant in position 4 ($dev4$) after hearing 3 standards is $1/3$. However, if the deviant is not located in position 4, it must be located in either position 5 or 6, which makes the likelihood of finding a deviant in position 5 ($dev5$) after hearing 4 standards $1/2$. The likelihood of finding a deviant in position 6 ($dev6$) after hearing 5 standards is $1$.

      To address the first research question, whether neural populations of human IC and MGB encode FM-rate and FM-direction, we tested whether these two nuclei show SSA to the FM-sweeps used in the experiment; namely, if neural responses in IC and MGB adapt to repeated FM-sweeps while preserving high responsiveness to FM-sweeps that deviate from the standards in FM-rate or FM-direction (Fig~\ref{fig:design}D). Since all sweeps were designed to elicit the same average activation across the tonotopic axis and elicited the same pitch percept, neural populations showing SSA to these FM-sweeps necessarily comprise neurons that are sensitive to FM-rate and FM-direction.

      To address the second research question, whether IC and MGB responses encode FM-sweeps as prediction error with respect to the listener expectations, we used Bayesian model comparison. We considered two models. The first model assumed that adaptation to repeated fast FM-sweeps was mostly driven by habituation to the stimulus sequence properties, independently of participant's expectations; namely, that neural populations habituate to repetitions of the standard, but show recovered responses to deviant irregardless of their position (habituation hypothesis; Fig~\ref{fig:design}E, h1). The second model assumed that adaptation was driven by predictive coding; namely, that neural responses to the deviants reflect prediction error with respect to the expectations of the participants (predictive coding hypothesis; Fig~\ref{fig:design}E, h2).

      We measured BOLD responses in participants' IC and MGB with an fMRI-sequence at 3-Tesla. The sequence was optimised to measure BOLD at relatively high spatial resolution (1.75\,mm isotropic) while maintaining a high SNR (around 25). 
      
    \subsection*{Behavioural Responses}

\begin{figure}[tbh!]
        \centering
        \includegraphics[width=0.6\textwidth]{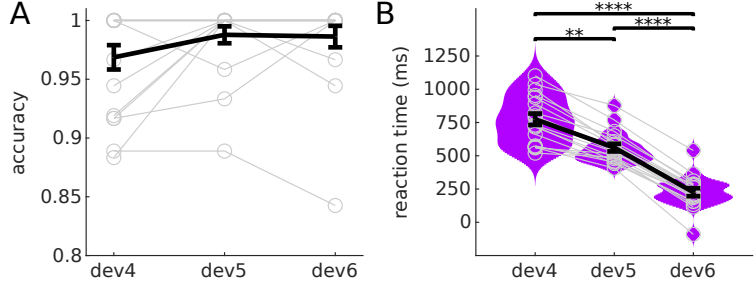}
        \vspace{1em}
        \caption{\textbf{Performance and reaction times}. Mean accuracy (A) and reaction times (B) across deviant positions. Grey circles represent the average value per participant and deviant position. Violin plots are the kernel density estimations of the reaction times for each deviant position. **: $p < 0.005$, **** $p < 0.00005$; all $p$-values corrected for 3 comparisons.
        \label{fig:behavioural}}
      \end{figure}

      All participants showed accuracies over 0.96 to all deviant positions (Fig~\ref{fig:behavioural}A). Accuracy was slightly higher for the two more expected deviant positions, but differences between conditions were not significant ($p > 0.1$, uncorrected). Reaction times (Fig~\ref{fig:behavioural}B) showed a behavioural benefit of expectations: Participants reacted faster to more expected deviants (average $RT = 770$\,ms, $558$\,ms and $246$\,ms for deviants at positions 4, 5, and 6, respectively (Fig~\ref{fig:behavioural}C); all differences were significant with $p < 0.0001$, corrected for 3 comparisons).
      
    \subsection*{Human IC and MGB show stimulus specific adaptation (SSA) to FM-sweeps}

      We first studied whether the IC and MGB show SSA to fast FM-sweeps to test if the two nuclei are sensitive to FM-rate and FM-direction in humans. We estimated BOLD responses using a general linear model (GLM) with 6 different regressors: the first standard ($std0$), the standards after the first standard but before the deviant ($std1$), the standards after the deviant ($std2$), and deviants at positions 4, 5, and 6 ($dev4$, $dev5$, and $dev6$, respectively; Fig~\ref{fig:design}D). The conditions $std1$ and $std2$ were parametrically modulated \cite{ODoherty2007} according to their positions to account for possible variations in the responses over subsequent repetitions (see Methods and Fig~\ref{fig:glmdesign}). 

\begin{figure}[tbh!]
        \centering
        \includegraphics[width=\textwidth]{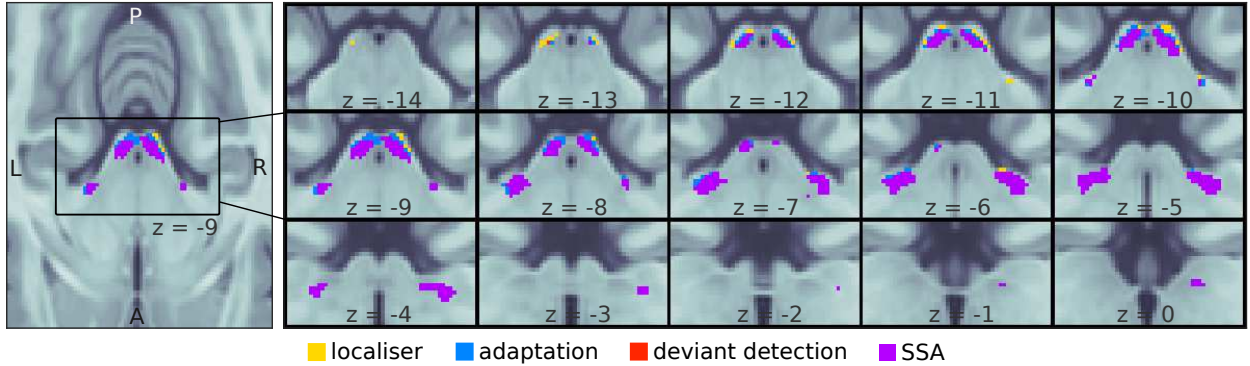}
        \vspace{1em}
        \caption{\textbf{Mesoscopic stimulus specific adaptation (SSA) in bilateral IC and MGB}. Regions within the MGB and IC ROIs adapted to the repeated standards (adaptation; blue shows adaptation only, purple shows SSA, which includes adaptation) and recovered responses to deviants (deviant detection; red shows deviant detection only, purple shows SSA, which includes deviant detection). Stimulus specific adaptation (i.e., recovered responses to a deviant in voxels showing adaptation; SSA) occurred in bilateral MGB and IC (purple). Maps were computed by thresholding the contrast $p$-maps at \mbox{$\text{\emph{FDR}} < 0.05$}. Yellow patches show voxels included in the anatomical masks computed with a functional localiser that showed neither adaptation nor deviant detection.
        \label{fig:ROIs}}
      \end{figure}

      To compute SSA, we determined which voxels within the ICs and MGBs adapted to the standard (i.e., adaptation) and recovered responsiveness to deviants (i.e., deviant detection). SSA regions were then defined as the intersection between adaptation and deviant detection regions. ICs and MGBs were identified based on structural MRI data and an independent functional localiser (see Methods; IC and MGB ROIs; coloured patches in Fig~\ref{fig:ROIs}). Within these ROIs, we used non-parametric ranksum tests ($N = 18$; one sample per participant) to find which voxels showed significant adaptation to repeated standards (contrast $std0 > 0.5\,std1 + 0.5\,std2$). The associated $p$-maps were thresholded so that the false-discovery-rate $\text{\emph{FDR}} < 0.05$. Surviving voxels constituted the \emph{adaptation} ROIs (blue and purple patches in Fig~\ref{fig:ROIs}). The same procedure was used to delimit the \emph{deviant detection} ROIs (red and purple patches in Fig~\ref{fig:ROIs}): the set of voxels within each anatomical ROI that responded significantly stronger to deviants than to repeated standards (contrast $dev4 > 0.5\,std1 + 0.5\,std2$; note that we compare the responses to the repeated standards with $dev4$ as this is the deviant position for which participants have the lowest expectation). The four anatomical ROIs showed significant adaptation (peak $p\leq 0.0001$) and deviant detection (peak $p < 0.0001$; cluster size, exact peak $p$-values and MNI coordinates are shown in Table~\ref{tab:roiStats}; all $p$-values corrected for four comparisons).

      SSA regions were computed combining the unthresholded \emph{adaptation} and \emph{deviant detection} $p$-maps. The uncorrected $p$-value associated to SSA for a given voxel was $p_{\text{SSA}} = \max{(p_{\text{adaptation}}, p_{\text{deviant detection}})}$. SSA $p$-maps where thresholded to $FDR < 0.05$ to compute the \emph{SSA} ROIs (Fig~\ref{fig:ROIs}, purple). The four anatomical ROIs had extensive SSA regions (cluster sizes larger than 90\,$mm^3$; peak $p \leq 0.0003$; exact peak $p$-values and MNI coordinates are shown in Table~\ref{tab:roiStats}; all $p$-values corrected for four comparisons). 

      \begin{table}[tbh!]
        \begin{tabularx}{\textwidth}{rcccccc}
          contrast             &    ROI     & cluster size & MNI coordinates (mm) &  peak-level $p$-value      \\
          \hline
          adaptation           & left IC    &  130 voxels  &   $[ -4,-35, -9]$    &     $p = 1 \times 10^{-4}$ \\
                               & right IC   &  124 voxels  &   $[  4,-35, -9]$    &     $p = 8 \times 10^{-5}$ \\
                               & left MGB   &  152 voxels  &   $[-14,-25, -7]$    &     $p = 8 \times 10^{-5}$ \\
                               & right MGB  &  146 voxels  &   $[ 14,-26, -6]$    &     $p = 1 \times 10^{-4}$ \\
          \\
          \hline
          deviant detection    & left IC    &   92 voxels  &   $[ -6,-33,-10]$    &     $p = 9 \times 10^{-5}$ \\
                               & right IC   &   91 voxels  &   $[  6,-33, -8]$    &     $p = 7 \times 10^{-5}$ \\
                               & left MGB   &  136 voxels  &   $[-14,-24, -7]$    &     $p = 5 \times 10^{-5}$ \\
                               & right MGB  &  140 voxels  &   $[ 11,-27, -5]$    &     $p = 2 \times 10^{-5}$ \\
          \\
          \hline
          SSA                  & left IC    &   91 voxels  &   $[ -4,-35, -9]$    &     $p = 3 \times 10^{-4}$ \\
                               & right IC   &   91 voxels  &   $[  6,-33, -9]$    &     $p = 2 \times 10^{-4}$ \\
                               & left MGB   &  136 voxels  &   $[-14,-25, -7]$    &     $p = 2 \times 10^{-4}$ \\
                               & right MGB  &  140 voxels  &   $[ 12,-26, -5]$    &     $p = 1 \times 10^{-4}$ \\
          \hline
        \end{tabularx}
        \vspace{1em}
        \caption{\textbf{Statistics and MNI coordinates of the adaptation and deviant detection contrasts in the four regions of interest.} All $p$-values FDR-corrected for the number of voxels in each anatomical ROI and further corrected for 4 comparisons within each contrast.
        \label{tab:roiStats}}
      \end{table}

      Significant SSA was also found at the single-subject level in 15 of the 18 participants ($p\leq0.048$ for each of the 15 participants, corrected for the 596 voxels included in a global subcortical auditory ROI that comprised bilateral IC and MGB), but not all participants showed significant SSA in all ROIs (IC-L: 8 participants, $p\leq0.049$; IC-R, MGB-L, MGB-R: 6 participants each, with $p\leq0.048$; all $p$-values corrected for the number of voxels in the ROI and further corrected for four ROIs).

    \subsection*{Human IC and MGB are sensitive to FM-direction and FM-rate}
    
      In the next step, we specifically tested whether the IC and MGB are similarly sensitive to FM-rate and FM-direction. To do that we analysed the data corresponding to: 1) trials where the standard and deviant differed only in modulation direction but not in absolute modulation rate; and 2) trials where the standard and deviant differed only in modulation rate but not in direction. If IC and MGB encode direction and rate, we would expect similar results in both partitions of the data. Conversely, if human IC and MGB are only sensitive to one of the two properties, we would expect null effects in the partition of the data where the standard and deviants differ only in that property.
       
\begin{figure}[tbh!]
        \centering
        \includegraphics[width=\textwidth]{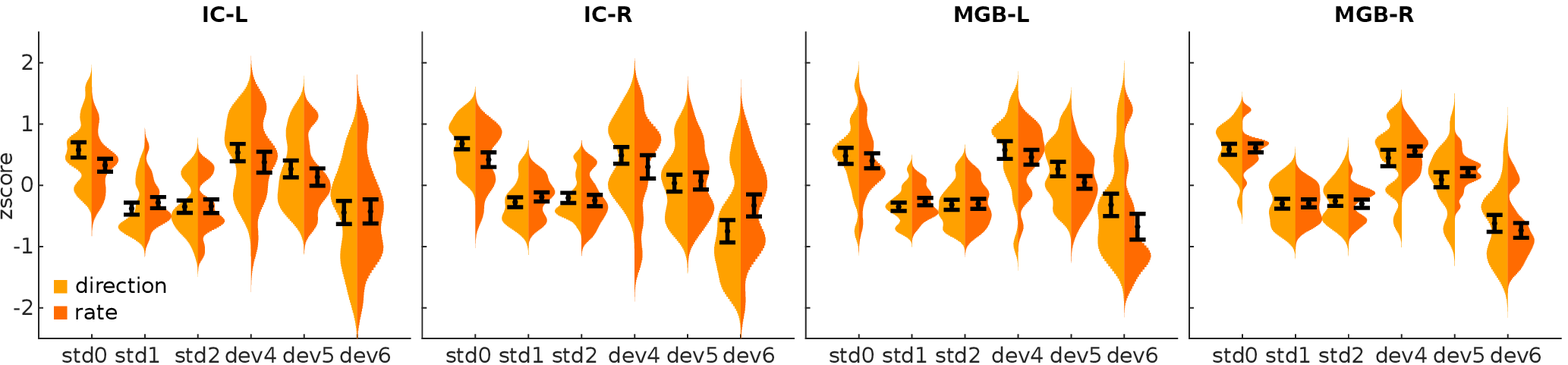}
        \vspace{1em}
        \caption{\textbf{Summary BOLD responses for partitions of the data where deviant and standard differed only in direction or rate.} Average z-score in each of the four SSA ROIs to the different experimental conditions in trials where the standard and deviant differed only in direction (orange) or rate (yellow). Violin plots are kernel density estimations of the distribution of $z$-scores, averaged over voxels and runs of each ROI. Each distribution holds 17 samples, one per participant (one participant was excluded from this analysis because there were not enough trials available, see Methods for details). Black error bars show the mean and standard error of the distributions.
        \label{fig:rate}}
      \end{figure}

      Results were similar in both partitions of the data (Fig~\ref{fig:rate}), demonstrating that the human IC and MGB encode both FM-direction and FM-rate. 

      We further corroborated that the levels of SSA were comparable for both types of FM changes at the single-subject level. In order to characterise FM-sensitivity with a number for each subject and FM-sweep combination, we used the SSA index \cite{Ulanovsky2003} $SI$ (Eq~\eqref{eq:SSAi}; note that $SI > 0$ is equivalent to the deviant detection contrast used in Fig~\ref{fig:ROIs}). 
       
      \begin{equation}
          SI = \frac{dev4 - \frac{1}{2}(std1 + std2)}{dev4 + \frac{1}{2}(std1 + std2)}
        \label{eq:SSAi}
      \end{equation}
    
      We measured the difference in $SI$ to FM-direction ($SI_{dir}$) and FM-rate ($SI_{rate}$) in the voxels of the subject-specific SSA regions calculated in the previous section for each of the 15 subjects for which we obtained significant SSA. If FM-direction and FM-rate are both encoded in IC and MGB, we would expect no difference between these two partitions of the data. We measured the difference using Cohen's $d = (\langle SI_{dir} \rangle - \langle SI_{rate} \rangle)/\sigma$), where $\langle SI \rangle$ is the average of $SI$ and $\sigma$ is the pooled standard deviation. The difference ranged between $d \geq -0.33$ and $d\leq 0.475$ across participants. The expected value of the difference ($E[d] = 0.02\pm0.05$) overlapped with zero, indicating once again that both FM-direction and FM-rate are already encoded in the subcortical auditory pathway. 

    \subsection*{Expectations drive the encoding of FM-sweeps in IC and MGB}

\begin{figure}[tbh!]
        \centering
        \includegraphics[width=\textwidth]{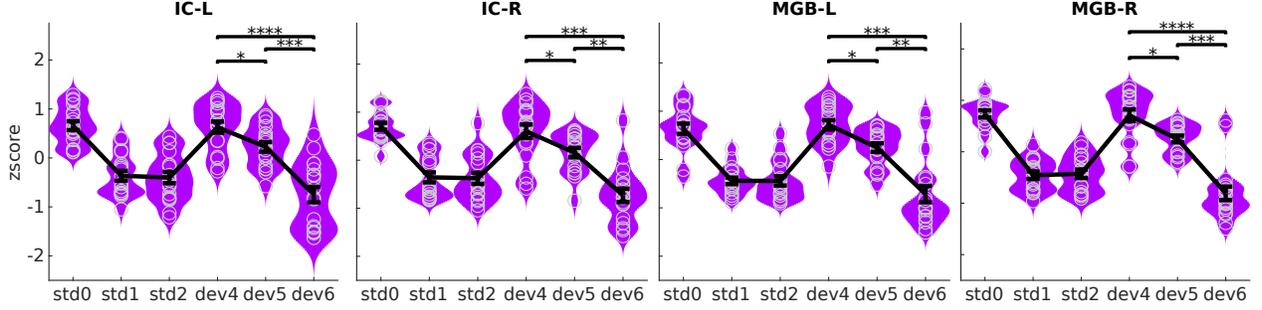}
        \vspace{1em}
        \caption{\textbf{Summary BOLD responses.} Average z-score in each of the four SSA ROIs to the different regressors. Violin plots are kernel density estimations of the distribution of $z$-scores, averaged over voxels and runs of each ROI. Each distribution holds 18 samples, one per participant. Black error bars show the mean and standard error of the distributions. Significance bars were computed by pooling across standard-deviant combinations. $Std0$, first standard; $std1$: standards preceding the deviant; $std2$: standards following the deviant; $dev4$, $dev5$ and $dev6$: deviants at positions 4, 5, and 6, respectively (Fig~\ref{fig:design}D). * $p < 0.05$, ** $p <0.005$, *** $p < 0.0005$, **** $p < 0.00005$; all $p$-values corrected for 12 comparisons.
        \label{fig:betas}}
      \end{figure}

      To address our second question, we evaluated whether the average BOLD responses to deviants in the three different positions were affected by participant's subjective expectations within the SSA regions. In congruence with the predictive coding hypothesis (Fig~\ref{fig:design}E, h2), the response profile showed reduced responses for more expected deviants (Fig~\ref{fig:betas}).

      Formal statistical testing confirmed that responses to different deviant positions were different in all ROIs for all contrasts among deviant positions: $dev4 \neq dev5$ (\mbox{$|d| \geq 0.99$} and \mbox{$p < 0.006$}), $dev4 \neq dev6$ (\mbox{$|d| \geq 2.39$} and \mbox{$p < 0.00005$}), and $dev5 \neq dev6$ (\mbox{$|d| \geq 1.74$} and \mbox{$p < 0.0003$}; all $p$-values corrected for $3\times4$ 12 comparisons). Exact $p$-values and effect sizes are listed in Table~\ref{tab:betas}. All statistical tests included one sample per participant, ROI, and deviant position.

      \begin{table}[tbh!]
        \centering
        \begin{tabularx}{0.7\textwidth}{rcccc}
               &          \multicolumn{4}{c}{\textbf{IC-L}}          \\
          \hline
               &        \multicolumn{2}{c}{dev5}        &       \multicolumn{2}{c}{dev6}    \\
          dev4 & $d = 0.89$  & $p = 0.017$           & $d = 2.35$  & $p = 4\times10^{-5}$    \\
          dev5 &        \multicolumn{2}{c}{  }        & $d = 1.75$  & $p = 4\times10^{-4}$    \\
          \\
               &          \multicolumn{4}{c}{\textbf{IC-R}}          \\
          \hline
               &        \multicolumn{2}{c}{dev5}        &       \multicolumn{2}{c}{dev6}    \\
          dev4 & $d = 0.86$  & $p = 0.022$           & $d = 2.24$  & $p = 10^{-4}$    \\
          dev5 &        \multicolumn{2}{c}{  }        & $d = 1.74$  & $p = 5\times10^{-4}$    \\
          \\
               &          \multicolumn{4}{c}{\textbf{MGB-L}}          \\
          \hline
               &        \multicolumn{2}{c}{dev5}        &       \multicolumn{2}{c}{dev6}    \\
          dev4 & $d = 1.20$  & $p = 0.0075$           & $d = 2.46$  & $p = 10^{-4}$    \\
          dev5 &        \multicolumn{2}{c}{  }        & $d = 1.68$  & $p = 0.0015$    \\
          \\
               &          \multicolumn{4}{c}{\textbf{MGB-R}}          \\
          \hline
               &        \multicolumn{2}{c}{dev5}        &       \multicolumn{2}{c}{dev6}    \\
          dev4 & $d = 1.17$  & $p = 0.0073$           & $d = 2.91$  & $p = 2\times10^{-5}$    \\
          dev5 &        \multicolumn{2}{c}{  }        & $d = 2.40$  & $p = 3\times10^{-4}$    \\
          \\
          \hline
        \end{tabularx}
        \vspace{1em}
        \caption{\textbf{Statistics of the average BOLD response differences between deviant positions.} Effect size is expressed as Cohen's $d$. Statistical significance was evaluated with two-tailed Ranksum tests between the distributions of the mean response in each ROI across participants ($N = 18$), pooling across standard-deviant combinations. All $p$-values in the table are corrected for $3\times4 = 12$ comparisons. 
        \label{tab:betas}}
      \end{table}

      To corroborate that differences were present at the single-subject level we run a correlation analysis for each of the 15 participants for which we obtained significant SSA. In each participant, we computed the Pearson's correlation between the BOLD responses elicited by each deviant location with its likelihood of occurrence (namely, $1/3$ for deviant 4, $1/2$ for deviant 5, and $1$ for deviant 6). If BOLD responses reflect prediction error, we would expect a negative correlation between the likelihood and the responses. We found significantly negative correlations in all 15 participants ($\rho \in [-0.87, -0.42]$, all $p < 0.03$; all Pearson tests had $9\times3 = 27$ samples, 3 per run).
          
    \subsection*{FM-sweeps are encoded as prediction error in the IC and MGB}

      We used Bayesian model comparison to formally evaluate the response properties in each voxel of the IC and MGB ROIs. This approach provides for a quantitative assessment of the likelihood that each of the two hypotheses (Fig~\ref{fig:design}E) can explain the responses in each voxel. This analysis is sensitive to possible region-specific effects that could have been averaged out when aggregating the z-scores across voxels in each ROI.
          
\begin{figure}[tbh!]
        \centering
        \includegraphics[width=\textwidth]{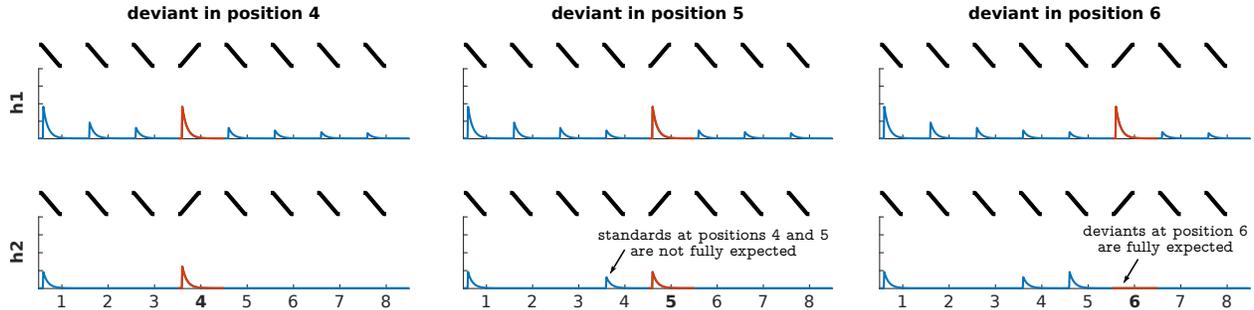}
        \vspace{1em}
        \caption{\textbf{Design of the Bayesian models.} Each model was defined according to the relative amplitudes it predicted for the different sounds in the sequences. H1 assumed asymptotic habituation to consecutive standards and recovered responses to deviants. H2 assumed that responses to the stimuli depended on how predictable they were.
        \label{fig:bayesdes}}
      \end{figure}

      Following the methodology described in \cite{Rosa2010, Stephan2009}, we first calculated the log-likelihood of each model in each voxel of the two ICs and MGBs in each participant. Each model yields different predictions on the relative amplitudes to different positions in the sequences (Fig~\ref{fig:design}E). We tested h1 and h2 to adjudicate between the habituation and predictive coding explanations of the responses. H1 assumed an asymptotic decay of the standards and recovered responses to the deviants; h2 assumed that the responses to both deviants and standards would depend on the participant's expectations (Fig~\ref{fig:bayesdes}; for exact values, see Methods). Participant-specific log-likelihoods were used to compute the Bayes factor $K$ (i.e., the ratio of the posterior likelihoods) between h1 and h2.

\begin{figure}[tbh!]
        \centering
        \includegraphics[width=0.35\textwidth]{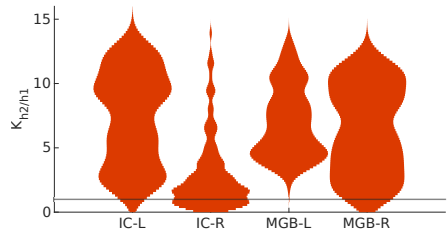}
        \vspace{1em}
        \caption{\textbf{Bayesian model comparison.} Bayes' factor $K$ for the model comparison h2/h1. Values $K > 1$ are more favourable to h2; values $K < 1$ mean that h1 is a better explanation of the data than h2. $K = 1$ is indicated by the fine grey line. Violin plots are kernel-density estimations of the distribution of $K$ across voxels (i.e., one sample per voxel).
        \label{fig:bayesDist}}
      \end{figure}

\begin{figure}[tbh!]
        \centering
        \includegraphics[width=\textwidth]{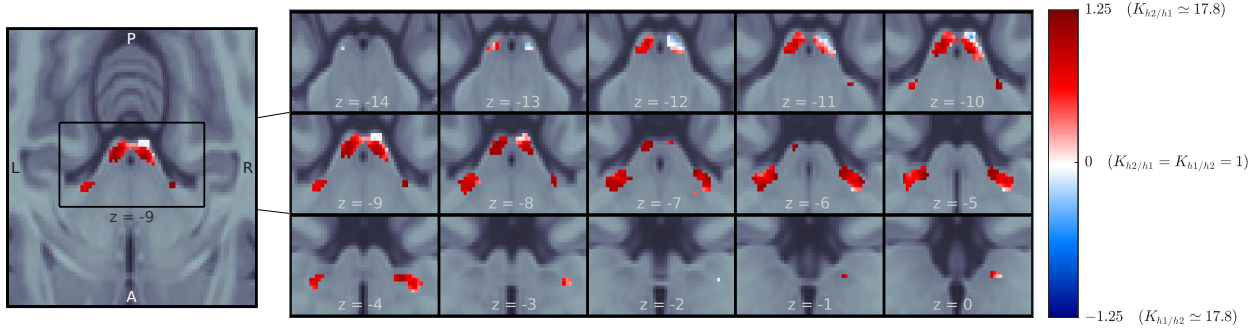}
        \vspace{1em}
        \caption{\textbf{Topographic distribution of each model.} Bayes' factor $K$ between h2 (predictive coding) and h1 (habituation) in each of the voxels of the subcortical ROIs in a logarithmic scale. Voxels with negative $\log K$ values ($K < 1$; blue) are best explained by h1; voxels with positive $\log K$ values ($K > 1$; red) are best explained by h2.
        \label{fig:bayesMap}}
      \end{figure}

      H2 was the best explanation for the data in the majority of voxels of the four ROIs (Figures~\ref{fig:bayesDist} and~\ref{fig:bayesMap}): h2 was more likely than h1 in 99\% and 80\% of the left and right IC, respectively, and in all voxels of the left and right MGB.

      To test whether the effect was present at the single-subject level, we computed $K$ independently for each subject in the subject-specific bilateral IC and MGB. Since we performed the group analyses over the entire ROIs (and not only the SSA regions), here we used the full anatomical ROIs of each participant. We measured for how many voxels within each region and participant h2 was the better explanation of the data. H2 was the better explanation of the data (more than 50\% of voxels) in 16 (IC-L), 13 (IC-R), 13 (MGB-L), and 15 (MGB-R) participants. In 7 (IC-L), 5 (IC-R), 10 (MGB-L), and 8 (MGB-R) participants H2 was the better explanation even in more than 75\% of voxels.
          
    \subsection*{FM-sweeps are encoded as prediction error in primary and secondary MGB}

      The auditory pathway is anatomically subdivided into two sections: the primary (lemniscal) or secondary (non-lemniscal) pathways. The primary pathway is characterised by neurons that carry auditory information with high fidelity and it is generally regarded as responsible for the transmission of bottom-up sensory input \cite{Hu2003}. The secondary pathway has wider tuning curves and it is generally regarded as responsible for the integration of contextual and multisensory information \cite{Hu2003}. 
      
      Both IC and MGB comprise regions that participate in both, the primary and secondary pathways \cite{Hu2003}. The primary subdivision of the IC is its central nucleus, while the cortices constitute the secondary subdivisions. The primary subdivision of the MGB is its ventral section, while the medial and dorsal sections constitute the secondary subdivisions.
      
      In rodents, SSA and prediction error to pure tones are significantly stronger in secondary subdivisions (e.g.,~\cite{Parras2017}). In humans, prediction error is similarly strong in primary and secondary MGB for pure tones \cite{Tabas2020}. Here we test for differential representations of prediction error to FM-sweeps in MGB. 
      
\begin{figure}[tbh!]
        \centering
        \includegraphics[width=0.8\textwidth]{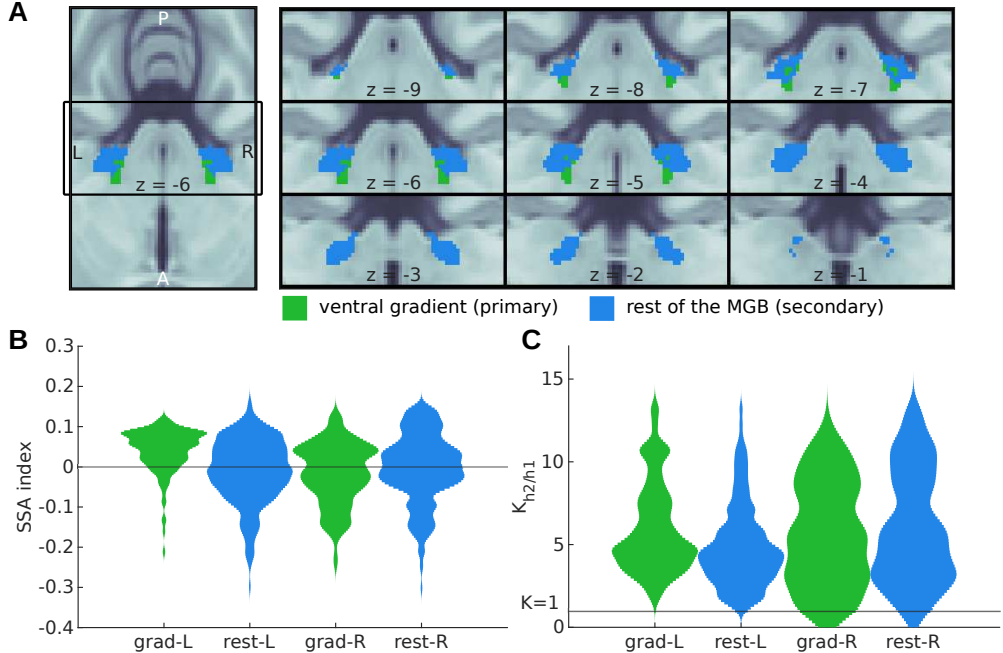}
        \vspace{1em}
        \caption{\textbf{Analyses of BOLD responses in ventral MGB.} A) Masks from~\cite{Mihai2019} of the ventral MGBs (green); blue indicates the remainder of the anatomical MGB ROIs. B) The distribution of the SSA index $SI$ across each of the two subdivisions of the MGB ROIs; $SI> 0$ is usually interpreted as SSA in the animal literature \cite{Ulanovsky2003}. C) Histograms showing Bayes' factor K for the comparison between h2 and h1 (Fig~\ref{fig:design}E) in each of the subdivisions. No systematic functional differences are apparent between primary and secondary MGB.
        \label{fig:subdivisions}}
      \end{figure}

      Distinguishing between the primary and secondary subsection of the IC and MGB non-invasively is technically challenging \cite{Moerel2015}. A recent study \cite{Mihai2019} distinguished two distinct tonotopic gradients of the MGB. The ventral tonotopic gradient was identified as the ventral or primary (vMGB) subsection of the MGB (see Fig~\ref{fig:subdivisions}A, green). Although the parcellation is based only on the topography of the tonotopic axes and their anatomical location, the region is the best approximation to-date of the vMGB in humans. No parcellation of the IC is available to-date.

      Both primary and secondary subdivisions of bilateral MGB showed SSA. SSA strength was measured in each voxel using the SSA index (Eq~\eqref{eq:SSAi}). Distributions of the $SI$ across the voxels of each of the subdivisions were comparable in both hemispheres (Fig~\ref{fig:subdivisions}B) demonstrating that SSA is not confined to nor stronger in the secondary MGB.
      
      Predictive coding (h2) was the best explanation for the responses of all voxels in the two subdivisions of the left MGB, and in 95\% and 97\% of the primary and secondary subdivisions of the right MGB, demonstrating that FM-sweeps are encoded as prediction error in both, primary and secondary subdivisions of bilateral MGB. Moreover, the distributions of the Bayes' factor $K$ between the predictive coding (h2) and adaptation (h1) hypotheses were comparable across subdivisions (Fig~\ref{fig:subdivisions}C).

    \subsection*{Prediction error to FM-sweeps and pure tones has similar topographic distributions in the IC}
    
      To study whether the same neural populations are in charge of encoding prediction error to FM-sweeps and pure tones, we compared the topographic distribution of the Bayes factor $K$ between the h2 and h1 in our data with the topographic distribution of the Bayes factor $K$ we obtained in a previous experiment, where we measured BOLD responses to the same experimental paradigm as here but using pure tones \cite{Tabas2020}. We computed the correlation between both $K$ across voxels of each of the four ROIs, as defined by the anatomical atlas from~\cite{Sitek2019}.
       
      Distribution of $K$ to both families of stimuli was strongly correlated across voxels of bilateral IC (left, $\rho = 0.47$, $p = 10^{-7}$; right, $\rho = 0.34$, $p = 10^{-4}$; $p$-values corrected for 4 comparisons), but not across voxels of the MGBs (left, $\rho = -0.11$, $p = 0.22$; right, $\rho = 0.15$, $p = 0.1$; uncorrected $p$-values).

  \section*{Discussion} 

    \paragraph{} 
          
      The effects of expectations on speech recognition are readily evident in our daily lives. However, the neural mechanisms underlying the integration of expectations in at early stages of the speech processing pipeline are poorly understood. Here we have investigated how fast FM-sweeps, the dynamic building blocks of speech sounds, are encoded in the human subcortical auditory pathway, and how the subjective expectations of the listener influence their processing. Our study provided four main findings: first, we showed that the human IC and MGB comprise FM-direction and FM-rate selective neuronal populations. Second, we showed that responses in IC and MGB were driven by subjective expectations of the participants, demonstrating that the IC and MGB are integrated in a global network of predictive coding. The findings were robust and present at the single-subject level, demonstrating the generalisation power of the result. Third, we showed that the expectations determined the responses to FM-sweeps in primary and secondary subdivisions of bilateral MGB. Last, we showed that the topographic distribution of neural populations encoding the FM-sweeps as prediction error was similar to that of pure tones in the IC. 
      
      Combined, our results provide first demonstration that the human IC and MGB are actively engaged in the predictive processing of the building blocks of speech, fast FM-sweeps. This confirms the long-standing hypothesis that predictive coding combines high-level expectations with the exquisite temporal properties of the subcortical auditory pathway to promote the encoding of low-level features of the speech signal \cite{Yildiz2013, Kriegstein2008}. This mechanism might be responsible for boosting encoding efficiency and aiding speech recognition.
          
    \paragraph{} 
      
      Neurons that respond selectively to FM-direction and FM-rate have been located in rodents in the IC \cite{Hage2003, Li2010, Geis2013}, MGB \cite{Lui2003, Kuo2012}, and auditory cortex \cite{Zhang2003, Ye2010, Trujillo2013, Issa2016}. In contrast, FM-selectivity has only been reported in humans in auditory cortex \cite{Okamoto2015} or higher-order areas of the cerebral cortex \cite{Hsieh2012, Joanisse2014}. Here, we have established that neural populations in the human IC and MGB show SSA to FM-direction and FM-rate. Since our FM-sweeps were matched in duration, pitch, and expected elicited activity along the tonotopic axis, this finding provides first evidence for FM-selectivity as early as in the IC in humans.
      
    \paragraph{} 

      Animal studies have extensively shown that the SSA index to pure tones in IC and MGB increases with increasing rarity and frequency difference of the deviant with respect to the standard \cite{Duque2016, Duque2015, Ayala2015b, Ayala2013, Zhao2011, Antunes2011, Antunes2010, Anderson2009, Malmierca2009}. These studies implicitly assume that sensory neurons form expectations based on the local statistics of the stimuli. Whether adaptation driven by such potential expectations is a true reflection of predictive coding is, however, still a matter of debate \cite{Malmierca2015, Carbajal2018}. Modelling studies have demonstrated that identical behaviours can be produced by synaptic fatigue without the need of expectations \cite{Eytan2003, Mill2011, Mill2012}. Manipulating expectations orthogonally to stimulus regularities is the only way to assess if prediction error is computed with respect to a global model of the sensory world \cite{Tabas2021}. 
          
      To-date, the only evidence (see~\cite{Tabas2021} for a review) that subcortical nuclei encode stimuli according to subjective expectations independently of stimulus regularities was provided by our previous study on pure tones in human IC and MGB \cite{Tabas2020}. Here we used fast FM-sweeps that were explicitly designed to elicit the same activation across the tonotopic axis \cite{Tabas2021b} to ensure that participants had to make use of FM-direction and FM-rate selective neurons to differentiate the deviant from the standards. The current findings demonstrate that the same principles apply to the encoding of dynamic FM-sweeps, even though FM is decoded much later than pure tone frequency in the auditory system \cite{Kuo2012}. 
          
      Our results also showed that the topographic distribution of voxels encoding pure tones and FM-sweeps according to the principles of predictive coding were highly correlated in the IC, but not in the MGB. This divergence might indicate a different functional role of the IC and the MGB with respect to both families of stimuli; however, it might also be caused by a greater variability in the anatomical location and orientation of the MGB across subjects \cite{Moerel2015} and should be considered with caution until replications are available.
   
    \paragraph{}  
          
      Although the expectations we induced in our participants had a relatively simple structure, they had a comparable level of abstraction as expectation stemming from grammatical rules \cite{Gunter2000}, semantic context \cite{Davis2011}, or familiarity with the speaker's style \cite{Regel2010}. An integrated inverted hierarchy could propagate linguistic predictions all the way to formant transitions \cite{Tabas2021, Friston2021, Kriegstein2008}, and use neural centres in the IC and MGB to compute prediction error with respect to these predictions.
          
      The expectations induced by our paradigm are most likely generated in the cerebral cortex. However, since we optimised our paradigm to study prediction error rather than the generation of expectations, we cannot test whether the subcortical responses we measured are driven or not by corticofugal projections. This possibility would be consistent with the massive corticofugal connections from cerebral cortex to MGB and IC \cite{Winer1984, Winer2005}, and with results from animal studies where the deactivation of unilateral auditory cortex \cite{Bauerle2011} or the TRN \cite{Yu2009} led to reduction of SSA in the ventral MGB (but also see contradictory findings in non-lemniscal MGB \cite{Antunes2011} and non-lemniscal IC \cite{Anderson2013}).

    \paragraph{} 
    
      Despite the fact that the MGB is at a higher processing stage than the IC, we found similar prevalence of predictive coding in both nuclei for FM-sweeps (Fig~\ref{fig:bayesDist}) as well as for pure tones \cite{Tabas2020}. These results contrast with a study in rodents concluding that the MGB encodes prediction error more strongly than the IC \cite{Parras2017}. We speculate that this fundamental difference between the human and rodent studies is caused by the introduction of abstract rules in our paradigm: if prediction error is computed with respect to high-level expectations, there is no reason for prediction error to vary across the hierarchy, since the same expectations are used to compute prediction error at all levels. Rodent studies use passive listening tasks where expectations are induced by repetition. Without an explicit high-level model, prediction error can only be computed with respect to implicit local models that monitor local stimulus statistics and that may vary in complexity across processing stages.
      
      Our study provides also novel insight into characteristics of SSA. Previous studies on subcortical SSA rested almost exclusively on pure tones \cite{Malmierca2015, Carbajal2018, Tabas2021}. Only three studies considered whether SSA generalised to other acoustic properties. Thomas et al. \cite{Thomas2012} reported SSA to FM-rate in the IC of the big brown bat; however, since big brown bats use FM for echolocation and the authors used stimuli in the rate range of echolocation signals, it was unclear whether this behaviour would generalise to auditory FM. Gao et al. \cite{Gao2014} measured SSA using ramped and damped broadband noises in the IC, demonstrating that neurons in the IC adapt to intensity modulation. Last, Duque et al. \cite{Duque2016} measured SSA to intensity, and showed that neurons in the IC do not adapt to nominal loudness. Our findings complement these results showing that the human IC and MGB adapt to fast FM characteristic of speech without loudness or spectral changes, and provides first evidence for SSA to acoustic properties other than pitch and loudness in the subcortical pathways.
    
    \paragraph{}  

      Nuclei in the auditory pathway are organised in primary (or lemniscal) and secondary (or non-lemniscal) subdivisions. The lemniscal division of the auditory pathway has narrowly tuned frequency responses and is considered as responsible for the transmission of bottom-up information; the non-lemniscal division presents wider tuned frequency responses, is involved in multisensory integration, and is more heavily targeted by corticofugal connections \cite{Hu2003}. Because of these properties, it has been previously suggested that prediction error may be encoded exclusively in secondary subdivisions of the IC and MGB \cite{Ayala2015b, Malmierca2015, Parras2017}. In agreement with this hypothesis, SSA is stronger in secondary subdivisions of the rodent's IC \cite{Perez2012, Gao2014, Duque2014, Ayala2015b, Ayala2018} and MGB \cite{Antunes2010, Antunes2011, Duque2014}.
      
      In contrast, our results indicated an apparent lack of specialisation across subdivisions of the MGB during the encoding of FM-sweeps. Namely, both primary and secondary MGB were similarly responsive to FM, and they both encoded FM as prediction error. Similar results were apparent in our previous study when we investigated the encoding of pure tones \cite{Tabas2020}. This lack of specialisation would fit with the idea that expectations are used in the subcortical pathways to aid encoding: to optimise the resources of the subcortical stations requires to make use of the the narrow receptive fields of the primary subdivisions \cite{Hu2003}. 
      
      The fundamental difference between our results and the findings in animals might stem from a number of reasons. First, our design involved an active task: lemniscal pathways might only be strongly modulated by predictions when they carry behaviourally relevant sensory information. Second, the modulation of the subcortical auditory pathway might be fundamentally different in humans compared to other mammals, as they have to accomplish processing of such complex and dynamic signals as speech. Last, given the strength of the SSA effects reported in this study, it is possible that regions with weak SSA might have been contaminated with signal stemming from areas with strong SSA due to smoothing and interpolation necessary for the analysis of fMRI data.

    \paragraph{} 
          
      Given the paramount role of predictions on speech perception \cite{Blank2018, Sohoglu2012, Davis2011, Davis2007}, atypical predictive coding in the subcortical sensory pathway could have profound repercussion at the cognitive level \cite{McFadyen2020, Diaz2012, Tabas2021}. For instance, developmental dyslexia, a disorder characterised by difficulties with processing speech sounds, has been attributed to altered adaption dynamics to stimulus regularities \cite{Perrachione2016, Ahissar2006, Chandrasekaran2009}, altered responses in the left MGB \cite{Diaz2012, Chandrasekaran2009}, and atypical left hemispheric cortico-thalamic pathways \cite{MullerAxt2017, Tschentscher2019}. Understanding the mechanisms underlying the predictive processing of low-level features of speech in subcortical sensory pathways is an essential prerequisite to understand dysfunction.

  \section*{Methods}


      This study was approved by the Ethics committee of the Technische Universt\"{a}t Dresden, Germany (ethics approval number EK 315062019). All listeners provided written informed consent and received monetary compensation for their participation.

    \subsection*{Participants}

      Eighteen German native speakers (12 female), aged 19 to 31 years (mean 24.6), participated in the study. None of them reported a history of psychiatric or neurological disorders, hearing difficulties, or current use of psychoactive medications. Normal hearing abilities were confirmed with pure tone audiometry (250\,Hz to 8000\,Hz); all participants had hearing threshold equal to or below 15\,dB SPL in the frequency range of the FM sweeps (1000\,Hz-3000\,Hz). Participants were also screened for dyslexia (German SLRT-II test \cite{Moll2014}, RST-ARR \cite{Ibrahimovic2013}, and rapid automatised naming (RAN) test of letters, numbers, objects, and colours \cite{Denckla1974}) and autism (Autism Spectrum Quotient; AQ \cite{Baron2001}). All scores were within the neurotypical range (SLRT: $\min(\max(PR_{\text{words}}, PR_{\text{pseudowords}})) = 21$, higher than the cut-off value of 16, following the same guidelines as \cite{Gutschmidt2020}; RST-ARR: all $PR \geq 31$, higher than the  cut-off value of 16; RAN: maximum of 3 errors and $RT < 36$\,seconds in each of the four categories; AQ: all participants $AQ \leq 31$, under or equal to the cut-off value of 32).

      Since we had no estimations of the possible sizes of the effects, we maximised our statistical power by recruiting as many participant as we could fit in the MRI measurement time allocated to the study. This number was fixed to twenty before we started data collection, but two participants dropped out of the study during data collection. We maximised the amount of data collected for participant to reduce random error to a minimum and maximise the likelihood of measuring effects at the single-subject level.

    \subsection*{Experimental paradigm}

      All sounds were 50\,ms long (including 5\,ms in/out ramps) sinusoidal FM sweeps. The frequency sweeps lasted for 40\,ms and were preceded and followed by 5\,ms long segments of constant frequency that overlapped with the in/our ramps. The sweeps were assembled in the frequency space to avoid discontinuities of the final waveforms.

      We used a total of three sweeps during the experiment: a \emph{fast up} sweep with starting frequency $f_0 = 1000$\,Hz and ending frequency $f_1 = 1200\,$Hz ($\Delta f = 200$\,Hz); a \emph{slow up} sweep with $f_0 = 1070$\,Hz and $f_1 = 1170$\,Hz ($\Delta f = 100$\,Hz), and a fast down sweep with $f_0 = 1280$\,Hz and $f_1 = 1080$\,Hz ($\Delta f = -200$\,Hz). The sweeps had different average frequencies to ensure that they elicited the same average spectral activity along the tonotopic axis and the same pitch percept (see~\cite{Tabas2021b} for details).

      From those 3 sweeps we constructed 6 standard-deviant frequency combinations that were used the same number of times across each run, so that all sweeps were used the same number of times as deviant and standards. Each sequence consisted of 7 repetitions of the standard sweep and a single instance of the deviant sweep. Sweeps were separated by 700\,ms inter-stimulus-intervals (ISI), the shortest possible ISI that allowed the participants to predict the fully expected deviant \cite{Tabas2020}, amounting to a total duration of 5300\,ms per sequence. 

      In each trial of the fMRI experiment, participants listened to one tone sequence and reported, \emph{as fast and accurately as possible} using a button box with three buttons, the position of the deviant (4, 5 or 6). The inter-trial-interval (ITI) was jittered so that deviants were separated by an average of 5 seconds, up to a maximum of 11 seconds, with a minimum ITI of 1500\,ms. We chose such ITI properties to maximise the efficiency of the response estimation of the deviants \cite{Friston1999} while keeping a sufficiently long ITI to ensure that the sequences belonging to separate trials were not confounded.

      All but one participant completed 9 runs of the main experiment across three sessions; participant 18 completed only 8 runs for technical reasons. Each run contained 6 blocks of 10 trials. The 10 trials in each block used one of the 6 possible sweep combinations, so that all the sequences within each block had the same standard and deviant. Thus, within a block only the position of the deviant was unknown, while the deviant's FM-direction and FM-rate were known. The order of the blocks within the experiment was randomised. The position of the deviant was pseudorandomised across all trials in each run so that each deviant position happened exactly 20 times per run but an unknown amount of times per block. This constraint allowed us to keep the same a priori probability for all deviant positions in each block. In addition, there were 23 silent gaps of 5300\,ms duration (i.e., null events of the same duration as the tone sequences) randomly located in each run \cite{Friston1999}. Each run lasted around 10 minutes, depending on the reaction times of the participant.

      Due to an undetected bug in the presentation code, information on the exact sweep combination used in each trial was unavailable for some runs. The bug affected the first three runs of participants 1, 2, 4, and 5; and the first six runs of participant 3. This information was not relevant for the analyses that aggregated the data across sweep combinations, and affected only the analyses of Fig~\ref{fig:rate}, where we excluded the affected runs of participants 1, 2, 4, and 5, and participant 3 altogether.
      
      We also run a functional localiser that was designed to activate the participant's IC and MGB. Each run of the functional localiser consisted on 20 blocks of 16 seconds and lasted for about 6.5 minutes. Ten of the blocks were silent; the remaining blocks consisted on presentations of 16 sounds of one second duration each. Sounds were taken from a collection of 85 natural sounds collected by~\cite{Moerel2015}. Participants were instructed to press a key when the same sound was repeated twice to ensure that they attended the sounds; behavioural data from the functional localiser was not used in the analysis. 

      Each session consisted on three runs of the main experiment, interspersed with two runs of the functional localiser. All runs were separated by breaks of a minimum of 1 minute to allow the participants rest. Fieldmaps and a whole-head EPI (see~\emph{Data acquisition}) were acquired between the third and fourth run. In the first session, we also measured an structural image before the fieldmaps. The first run of the first session was preceded by a \emph{practice run} of four randomly chosen trials to ensure the participants had understood the task. We acquired fMRI during the practice run in order to allow the participants to undertake the training with MRI-noise. 

    \subsection*{Data acquisition}

      MRI data were acquired using a Siemens Trio 3\,T scanner (Siemens Healthineers, Erlangen, Germany) with a 32-channel head coil. 

      Functional MRI data were acquired using echo planar imaging (EPI) sequences. We used  partial coverage with 24 slices. The volume was oriented in parallel to the superior temporal gyrus such that the slices encompassed the IC, the MGB, and the superior temporal gyrus. In addition, we acquired one volume of an additional whole-head EPI with the same parameters (including the FoV) and 84 slices during resting to aid the coregistration process (see~\emph{Data preprocessing}).

      The EPI sequence had the following acquisition parameters: \mbox{TR = 1900\,ms}, TE = 42\,ms, flip angle 66$^{\circ}$, matrix size $88\times88$, FoV 154\,mm$\times$154\,mm, voxel size 1.75\,mm isotropic, bandwidth per pixel $1386$\,Hz/px, and interleaved acquisition. During functional MRI data acquisition, cardiac signal was acquired using a scanner pulse oximeter (Siemens Healthineers, Erlangen, Germany).

      Structural images were recorded using an MPRAGE \cite{Brant1992} T1 protocol with 1\,mm isotropic resolution, TE = 1.95\,ms, TR = 1000\,ms, TI = 880\,ms, flip angle 1 = 8$^{\circ}$, \mbox{FoV = 256\,mm$\times$256\,mm}.

      Stimuli were presented using MATLAB (The Mathworks Inc., Natick, MA, USA) with the Psychophysics Toolbox extensions \cite{Brainard1997} and delivered through an Optoacoustics (Optoacoustics Ltd, Or Yehuda, Israel) amplifier and headphones equipped with active noise-cancellation. Loudness was adjusted independently for each participant to a comfortable level before starting the data acquisition.

    \subsection*{Data preprocessing}

      The preprocessing pipeline was coded in Nipype 1.5.0 \cite{Gorgolewski2011}, and carried out using tools of the Statistical Parametric Mapping toolbox, version 12; Freesurfer, version 6 \cite{Fischl2002}; the FMRIB Software Library, version 5 (FSL) \cite{Jenkinson2012}); and the Advanced Normalization Tools, version 2.3 (ANTS) \cite{Avants2011}. All data were coregistered to the Montreal Neurological Institute (MNI) MNI152 1\,mm isotropic symmetric template. 

      First, we realigned the functional runs. We used SPM's \emph{FieldMap Toolbox} to calculate the geometric distortions caused in the EPI images due to field inhomogeneities. Next, we used SPM's \emph{Realign and Unwarp} to perform motion and distortion correction on the functional data. Motion artefacts, recorded using SPM's ArtifactDetect, were later added to the design matrix (see~\emph{Estimation of the BOLD responses}).

      Next, we used Freesurfer's recon-all routine to calculate the boundaries between grey and white matter (these are necessary to register the functional data to the structural images) and ANTs to compute the transformation between the structural images and the MNI152 symmetric template.

      Last, we coregistered the functional data to the structural image with Freesurfer’s \emph{BBregister}, using the boundaries between grey and white matter of the structural data and the whole-brain EPI as an intermediate step. Data was analysed in the participant space, and then coregistred to the MNI152 template. Note that, since the resolution of the MNI space (1\,mm isotropic) was higher than the resolution of the functional data (1.75\,mm isotropic), the transformation resulted in a spatial oversampling.

      All the preprocessing parameters, including the smoothing kernel size, were fixed before we started fitting the general linear model (GLM) and remained unchanged during the subsequent steps of the data analysis.

      Physiological (heart rate) data was processed by the PhysIO Toolbox \cite{Kasper2017}, that computes the Fourier expansion of each component along time and adds the coefficients as covariates of no interests in the model's design matrix. 

    \subsection*{Estimation of the BOLD responses}

      First level analyses were coded in Nipype and carried out using SPM. Second level analyses were carried out using custom code in MATLAB. The coregistered data were first smoothed using a 2\,mm full-width half-maximum kernel Gaussian kernel with SPM's \emph{Smooth}. 

      The first level GLM's design matrix for the main experiment included 6 regressors: first standard (std0), standards before the deviant (std1), standards after the deviant (std2), and deviants in positions 4, 5, and 6 (dev4, dev5, and dev6, respectively; Fig~\ref{fig:design}). Conditions std1 and std2 were modelled using linear parametric modulation \cite{ODoherty2007}, whose linear factors were coded according to the position of the sound within the sequence to account for effects of habituation \cite{Tabas2020}. The first level GLM's design matrix for the functional localiser included 2 conditions: sound and silence. On top of the main regressors, the design matrix also included the physiological PhysIO and artefact regressors of no-interest.

    \subsection*{Definition of the anatomical and SSA ROIs}

      We used a recent anatomical atlas of the subcortical auditory pathway \cite{Sitek2019} to compute prior regions corresponding to the left IC, right IC, left MGB, and right MGB, respectively. The atlas comprises three different definitions of the ROIs calculated using 1) data from the big brain project, 2) postmortem data, and 3) fMRI in vivo-data. To compute the prior coarse region for each nuclei we combined the three masks and inflated the resulting regions with a Gaussian kernel of 1\,mm fwdh. 

      The final IC and MGB regions were computed by combining the prior coarse regions with the results from the contrast $sound > silence$ of the functional localiser. Within each region, we thresholded the contrast to increasingly higher values until the number of surviving voxels equalled the volume of the region reported in \cite{Sitek2019}; namely, 146 voxels for each of the ICs, and 152 for each of the MGBs.

      We used the coefficients of the GLM or beta estimates from the first level analysis to calculate the adaptation (Fig~\ref{fig:ROIs}, blue patches) and deviant detection (red patches) ROIs, defined as the sets of voxels within the IC and MGB ROIs that responded significantly to the contrasts $std0 > 0.5 std1 + 0.5 std2$ and $dev4 > 0.5 std1 + 0.5 std2$, respectively. Significance was defined as $p < 0.05$, family-wise-error (FDR)-corrected for the number of voxels within each of the IC/MGB ROIs. SSA voxels are defined as voxels that show both, adaptation and deviant detection; thus, we calculated an upper bound of the $p$-value maps for the SSA contrast as the maximum of the uncorrected $p$-values associated to the adaptation and deviant detection contrasts. The SSA ROIs (Fig~\ref{fig:ROIs}, purple patches) were calculated by FDR-correcting and thresholding the resulting $p$-maps at $\alpha = 0.05$. All calculations were performed using custom-made scripts (see~Data and code availability).

    \subsection*{Bayesian model comparison}

      The Bayesian analysis of the data consisted as well of first and second level analyses. In the first level, we used SPM via nipype to compute the log-evidence in each voxel of each participant for each of the four models (see Fig~\ref{fig:bayesdes}). The models were described using one regressor with parametric modulation whose coefficients corresponded to a simplified view of the expected responses according to each model (Table~\ref{tab:models}). The expected responses of each model were the same in all trials that had the same standard-deviant combination and deviant position. Given the model amplitude(s) $a_n$ and the timecourse of a voxel $y$, SPM calculates the log-evidence of the linear model $y = \sum \beta_n a_n + \xi$, where $\beta_n$ are the linear coefficients of each regressor and $\xi$ are noise terms.
      
      \begin{table}[tbh!]
        \centering
        \begin{tabularx}{0.8\textwidth}{rrcccccccc}
            \multicolumn{2}{l}{\textbf{h1}}      &   1   &   2   &   3   &   4   &   5   &   6   &   7   &   8   \\
            \hline 
                \multirow{3}{*}{$a_0$} & deviant at 4 & $1/1$ & $1/2$ & $1/3$ &  $1$  & $1/3$ & $1/4$ & $1/5$ & $1/6$ \\
                                     & deviant at 5 & $1/1$ & $1/2$ & $1/3$ & $1/4$ &  $1$  & $1/4$ & $1/5$ & $1/6$ \\
                                     & deviant at 6 & $1/1$ & $1/2$ & $1/3$ & $1/4$ & $1/5$ &  $1$  & $1/5$ & $1/6$ \\
                \hline
                \\
            \multicolumn{2}{l}{\textbf{h2}}      &   1   &   2   &   3   &   4   &   5   &   6   &   7   &   8   \\
            \hline 
                \multirow{3}{*}{$a_0$} & deviant at 4 & $1/2$ &  $0$  &  $0$  & $2/3$ &  $0$  &  $0$  &  $0$  &  $0$ \\
                                     & deviant at 5 & $1/2$ &  $0$  &  $0$  & $1/3$ & $1/2$ &  $0$  &  $0$  &  $0$ \\
                                     & deviant at 6 & $1/2$ &  $0$  &  $0$  & $1/3$ & $1/2$ &  $0$  &  $0$  &  $0$ \\
                \hline
        \end{tabularx}
        \vspace{1em}
        \caption{\textbf{Amplitudes of the models used for Bayesian Model Comparison}. H1 assumes an asymptotic decay ($a_0 \propto 1/n$ where $m$ is the position of the stimulus in the sequence) in the responses for all standards, a full response to deviants, and a recovery from the last standard before the deviant and the first standard after the deviant that is sufficient to make the responses to both standards comparable. H2 assumes that the responses scale with predictability ($a_0 = 1 - p$, where $p$ is the likelihood of finding the heard stimuli in each position). The models were defined exactly as in~\cite{Tabas2020}.
        \label{tab:models}}
      \end{table}
    
      Log-evidence maps for each participant were combined following the random-effects-equivalent procedure described in \cite{Rosa2010, Stephan2009} to compute the posterior probability maps associated to each model at the group level. We combined the maps using custom scripts (see~Data and code availability). Histograms shown in Figures~\ref{fig:bayesDist} and~\ref{fig:subdivisions} are kernel-density estimates computed with the distribution of the posterior probabilities across voxels for each of the SSA ROIs.

    \subsection*{Statistical analysis}

      All pairwise comparisons reported in the study were evaluated for significance using two-tailed Ranksum tests. Unless stated otherwise, $p$-values for all analyses that comprised multiple testing were corrected using the Holm-Bonferroni method. A result was deemed statistically significant when the corrected $p < 0.05$. 

    \subsection*{Data and code availability}

      Derivatives (beta maps and log-likelihood maps, computed with SPM) and all code used for data processing and analysis are publicly available in \url{https://osf.io/f5tsy/}.

  \bibliographystyle{ieeetr}
  \bibliography{bib}
  
  \newpage

  \section*{Supplementary materials}

    \renewcommand{\thefigure}{S\arabic{figure}}
    \setcounter{figure}{0} 

    \begin{figure}[hbtp]
     \centering
      \includegraphics[width=1\textwidth]{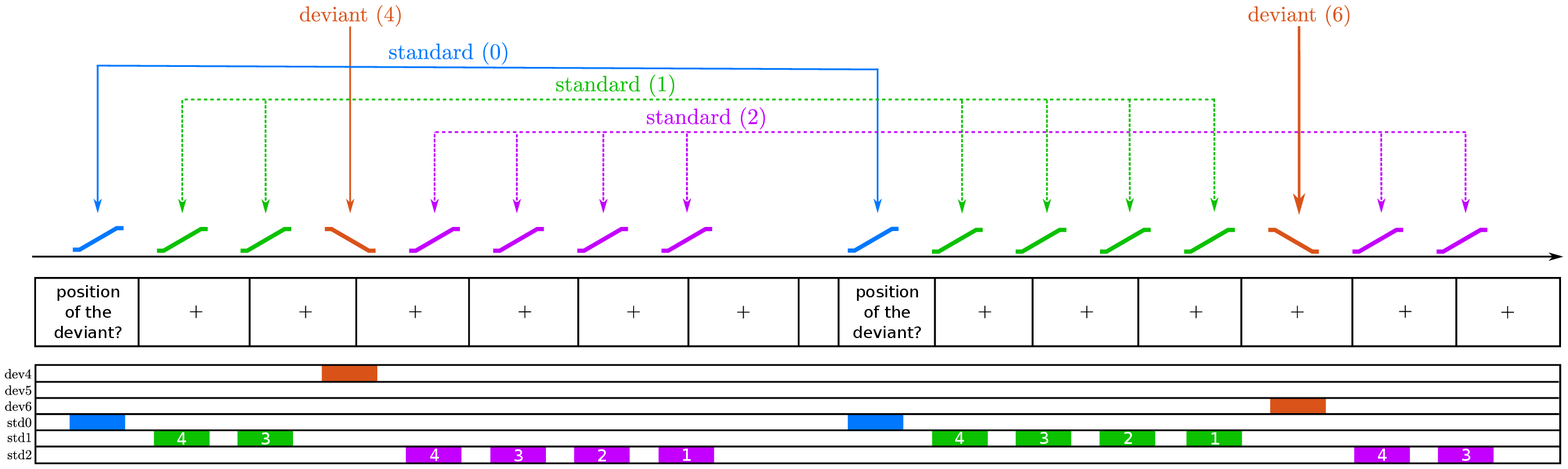}
      \caption{\textbf{Schematic of the GLM's design matrix.} An example of the GLM design matrix section corresponding to the regressors of interest prior to the convolution with the haemodynamic response function. The example includes two trials with two different deviant locations. The six regressors of interest were standard 0, standard 1, standard 2, deviant 4, deviant 5 (not shown), deviant 6. The standards were parametrically modulated and the modulation was equal to the inverted index of the standard within the sequence (i.e., 4 for the first $std1$, 3 for the second $std1$, etc; note that, since the modulation was mean-corrected before the fitting of the GLM, the absolute values of the modulation are not relevant). 
      \label{fig:glmdesign}}
    \end{figure}

\end{document}